\documentclass[aps,pre,reprint,floatfix,showpacs]{revtex4-1}
\usepackage{amsmath,amsfonts,amssymb}%
\usepackage{graphicx}%
\usepackage{hyperref}
\usepackage{placeins}
\usepackage{epstopdf}
\usepackage{multirow}

\graphicspath{{Figures/}{Figures/EPS/}}
\begin{document}

\title{Anisotropic diffusion of spherical particles in closely confining microchannels}

\author{Simon L Dettmer}
\author{Stefano Pagliara}
\author{Karolis Misiunas}
\author{Ulrich F Keyser}
\email{Corresponding author: ufk20@cam.ac.uk}
\affiliation{Cavendish Laboratory, University of Cambridge, 19 J J Thomson Avenue, Cambridge, CB3 0HE, United Kingdom}

\begin{abstract}
We present here the measurement of the diffusivity of spherical particles closely confined by narrow microchannels. Our experiments yield a two-dimensional map of the position-dependent diffusion coefficients parallel and perpendicular to the channel axis with a resolution down to $129~nm$. The diffusivity was measured simultaneously in the channel interior, the bulk reservoirs, as well as the channel entrance region. In the channel interior we found strongly anisotropic diffusion. While the perpendicular diffusion coefficient close to the confining walls decreased down to approximately $25\%$ of the value on the channel axis, the parallel diffusion coefficient remained constant throughout the entire channel width. In addition to the experiment, we performed finite element simulations for the diffusivity in the channel interior and found good agreement with the measurements. Our results reveal the distinctive influence of strong confinement on Brownian motion, which is of significance to microfluidics as well as quantitative models of facilitated membrane transport.
\end{abstract}

\maketitle

%%%%%%%%%%%%%%%%%%%%%%%%%%%%%%%%%%%%%%%%%%%%%%%%%%%%%%%%%%%%%%%%%%%%%%%%%%%%%%%%%%%%%%
%%%%%%%%%%%%%%%%%%%%%%%%%%%%%%%%%%%%%%%%%%%%%%%%%%%%%%%%%%%%%%%%%%%%%%%%%%%%%%%%%%%%%
\section{Introduction}
Diffusion in close confinement is paramount to transport across biological membranes and understanding the physical processes governing transport is of great relevance for designing drugs~\cite{Sugano2010}. Many molecules are transported across the membrane by passive diffusion through proteins that form long and narrow channels. Channel-facilitated diffusion has been studied experimentally~\cite{Benz1986,Hilty2001,Pagliara2013} as well as theoretically~\cite{Bezrukov2000,Berezhkovskii2002,Berezhkovskii2003,Berezhkovskii2005,Bezrukov2007,Bauer2006,Zilman2009,Zilman2010, Kolomeisky2007,Kolomeisky2011} and interpreting the models requires knowledge of the spatial dependence of diffusion coefficients inside the channel and at the entrance regions, either explicitly in the continuous models or implicitly in the form of diffusive hopping constants for discrete models. Besides the relevance to biological transport, it is also of interest in the study of physical phenomena such as entropic particle transport in corrugated channels for particle separation~\cite{Martens2011,Kettner2000}.
In the confinement of bounding walls, the diffusion coefficients of particles are decreased by viscous interactions with the walls as compared to the value in an infinite fluid. This hindered diffusion has been studied extensively for planar geometries involving spherical particles moving either above a single wall or between two plane walls~\cite{Benesch2003, Lobry1996,Brenner1961, Goldman1967,Lin2000,Sharma2010,Faucheux1994,Choi2007,Banerjee2005,Bevan2000,Leach2009,Michailidou2009,Ha2013}. 
However, to our knowledge, only one experimental study investigates position-dependent hindered diffusion in the presence of curved boundaries~\cite{Eral2010}. The authors studied the hindered diffusion of spherical particles inside closed cylinders that were considerably larger than the particles. Experiments on the diffusion of particles in closely confining channels~\cite{Cui2002} have been limited to effectively infinitely long channels and diffusion along the channel axis. So far, measurements of the position-dependent diffusion coefficients in closely-confining, finite length channels are lacking completely.
%%%%%%%%%%%%%%%%%%%%%%%%%%%%%%%%%%%%%%%%%%%%%%%%%%%%%%%%%%%%%%%%%%%%%%%%%%%%%%%%%%%%%%%%
%%%%%%%%%%%%%%%%%%%%%%%%%%%%%%%%%%%%%%%%%%%%%%%%%%%%%%%%%%%%%%%%%%%%%%%%%%%%%%%%%%%%%%%%%%%
\section{Methods}
\begin{figure}[hbtp]
	\centering
		\includegraphics[width=0.50\textwidth]{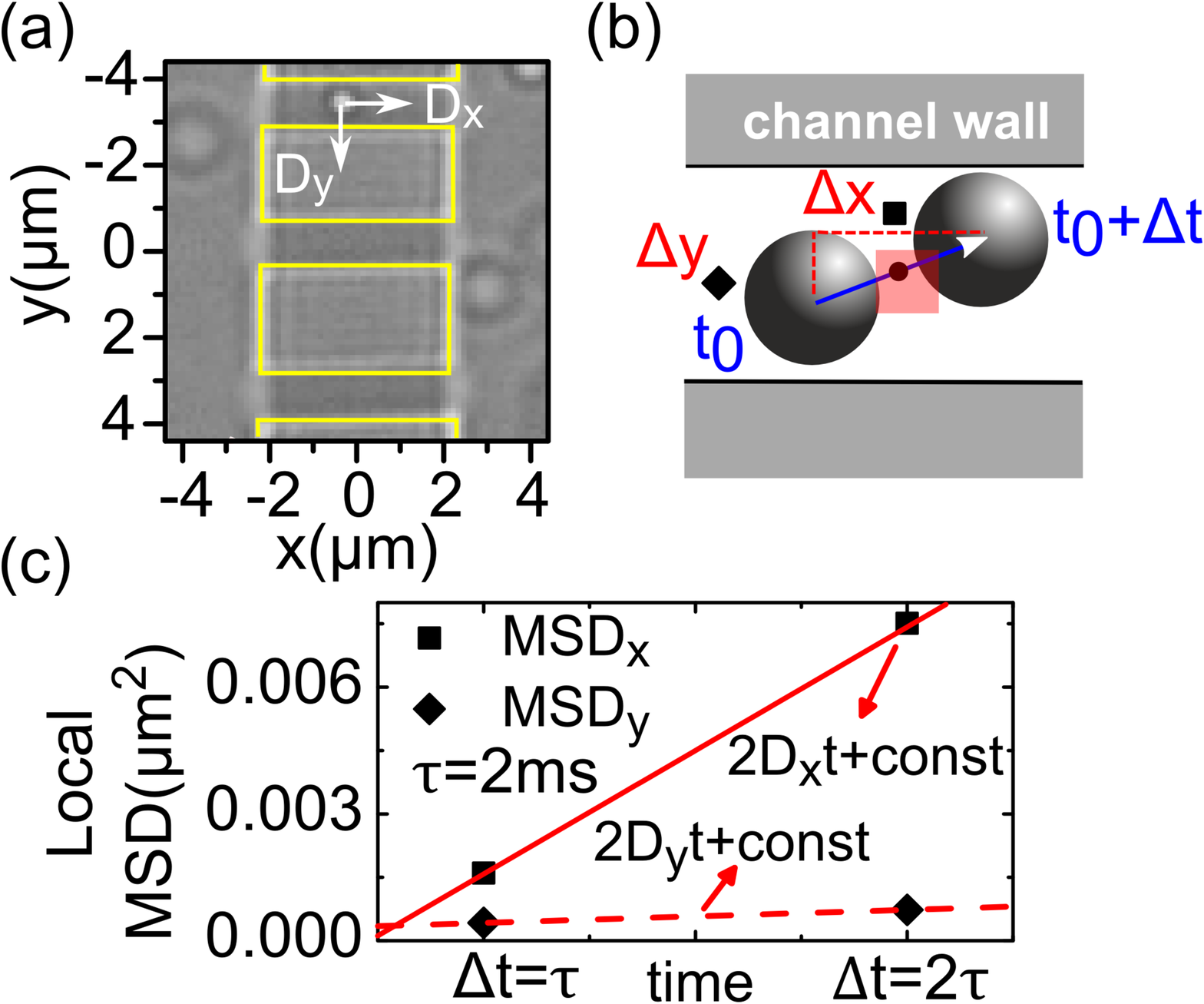}
	\caption{(Color online) Method for measuring local diffusion coefficients. (a) Particles diffusing in the microfluidic chip, containing two bulk reservoirs connected by three parallel microfluidic channels, are imaged by bright-field video microscopy. The channel edges are marked by the lines. We choose the coordinates such that the diffusion coefficients parallel and perpendicular to the channel walls are $D_x$ and $D_y$ respectively. (b) Displacements of tracked particles after time lags of one and two video frames are assigned to the bin of the midpoint of the displacement vector (marked by the box). (c) This yields the first two points of the $MSD$-vs-$t$-curves in both $x$ and $y$ directions for each bin. The slope of the linear fit yields the diffusion coefficients in $x$ and $y$, respectively.}
	\label{fig:Figure1}
\end{figure}
In this article we report the measurement of a complete two-dimensional~(2D) map with 129-nm resolution of the position-dependent diffusion coefficients of spherical particles. The polystyrene spheres [Polysciences (Warrington, PA); $505\pm 8$~nm in diameter] moved through an array of parallel, closely confining microchannels of semi-elliptical cross section (approximately 5~$\mu m$ in length and 1~$\mu m$ in width and height) separating two bulk reservoirs~[Figure~\ref{fig:Figure1}(a)]. Our data cover the channel interior as well as the entrance regions and the bulk reservoirs. The channels were realized in a microfluidic chip made in polydimethylsiloxane~(PDMS) via replica molding~\cite{Pagliara2011, Pagliara2013}. Briefly, for creating the mold an array of platinum wires was deposited on a silicon substrate via a focused ion beam. The wire cross section was measured \textit{in situ} by slicing the wire at one end, tilting the sample at 63$^\circ$, and imaging with a scanning electron microscope. Conventional photolithography, replica molding, and PDMS bonding to a glass slide were carried out to define 16-$\mu m$-thick reservoirs separated by a PDMS barrier and connected by the array of channels obtained as a negative replica of the platinum wires. The chip was filled with the particles dispersed in a 5~mM KCl solution and continuously imaged through an oil immersion objective (100$\times$ , 1.4 numerical aperture; UPLSAPO, Olympus). Illumination was provided from above by a light-emitting diode (Thorlabs MWLED). The transmitted light was collected by the objective and coupled to a complementary metal-oxide semiconductor camera (with a frame rate of 500~fps and a magnification of $\sim 8~pixels/\mu m$). With the objective having a depth of focus of approximately $2~\mu m$ and the focal plane close to the glass cover slide, particles were always tracked in proximity to at least one bounding wall.
Experiments were automated using a custom-made Lab$_{\text{VIEW}}$ program for positioning and video acquisition~\cite{Pagliara2013}. The temperature inside the chip during the experiment was monitored using a digital thermometer (RS Components, K-type thermocouple, 0.2\% accuracy).
Particle trajectories in two dimensions were extracted from the microscopy videos via a custom-written automated tracking algorithm with accuracies better than 20~nm inside the channels. Local diffusion coefficients were determined from a linear fit to local mean squared displacement~(\textit{MSD})-versus-time curves~\cite{DettmerRSI2014} [Figs.~\ref{fig:Figure1}(b) and \ref{fig:Figure1}(c)]. In short, we followed particles and measured their displacements for time lags of one and two frames, respectively. These displacements were assigned to the position bin of the mid point of the displacement vector. Subsequently, in each bin the measured values were averaged to give the first two points of local \textit{MSD}-versus-time curves for both $x$ and $y$ direction [Fig.~\ref{fig:Figure1}(c)]. The slope of the linear curve connecting these two points yields the diffusion coefficients. For creating the 2D map we binned the $xy$ positions into single camera pixel bins~(${129\times 129~nm^2}$). Trajectories of particles exploring the channels and the bulk were recorded for 80~min of video, corresponding to $2.453$ million frames and $5.506$ million tracked particle positions. \newline
For our numerical simulations we used $\text{COMSOL}$ Multiphysics 4.3b with the creeping flow module to solve the Stokes equation by the finite element method with an adaptive mesh size. The system treated was a spherical particle moving in an infinitely long channel of semi elliptical cross section. The particle was positioned on different grid points in the cross-sectional $yz$ plane and the viscous friction tensor $\boldsymbol{\nu}$ calculated for each position. We imposed no-slip boundary conditions on the sphere surface and the channel walls. Furthermore, we utilized a common computational approach~\cite{Happel1983} and switched to the frame of reference of the particle. Thereby the walls become moving, which is mathematically treated as a slip velocity on the wall surface ($\vec{v}=v \vec{e_x}$ or $\vec{v}=v \vec{e_y}$, corresponding to parallel and perpendicular diffusion). To compare the simulations with the measured diffusion coefficients in the channel interior we used the Stokes-Einstein relation~\cite{Einstein1905} which gave the ratios $D_x(y,z)/D_0$ and $D_y(y,z)/D_0$ (see Appendix~\ref{sec:Stokes-Einstein} for details).
For calculating the bulk diffusivity, $D_0$, we inserted the temperature measured inside the chip during the experiment, $T=301.7~K$, the particle radius $a=250~nm$ and the viscosity of water, $\eta (T)$~\cite{Kestin1978}, into the Stokes-Einstein equation, giving $D_0=1.08~\mu m^2/s$.
For arriving at the perpendicular dependence of diffusivity, $D(y)$, we averaged the values $D(y,z)$ over the entire $z$ range by random sampling in order to avoid mesh artifacts. In the experiments, our measured diffusion coefficients represent as well values averaged over the entire $z$ range of 
700~nm accessible to the particles. It is important to note that the strong confinement in $z$ direction combined with the semi elliptical cross section leads to suppressed axial position fluctuations inside the channels.\newline
The PDMS channel width was determined optically from the microscopy videos as well as from considering the width over which particles were tracked inside the channel. The widths of both measurement methods agreed and we found values of $(1.15\pm 0.13)~\mu m$ for the bottom two channels and $(1.02\pm 0.13)~\mu m$ for the top channel; thus all three channels had the same width within measurement accuracy. For the numerical simulations we used a width of $1.2~\mu m$ and assumed that the semi elliptical cross section of the platinum wires was preserved.
%%%%%%%%%%%%%%%%%%%%%%%%%%%%%%%%%%%%%%%%%%%%%%%%%%%%%%%%%%%%%%%%%%%%%%%%%%%%
%%%%%%%%%%%%%%%%%%%%%%%%%%%%%%%%%%%%%%%%%%%%%%%%%%%%%%%%%%%%%%%%%%%%%%%%%%%
\section{Results}
\FloatBarrier
\subsection{Dependence of the diffusion coefficient on the axial position}
\subsubsection{Diffusion coefficient parallel to the channel axis}
We first consider diffusion parallel to the channel axis ($D_x$). A two-dimensional color map of $D_x$ is shown in Fig.~\ref{fig:Diffusion_Dx_and_Dy}(a). We found a significantly reduced diffusivity inside the channels as compared to the values measured in the bulk. To quantify this further, we measured the dependence of the diffusion coefficient along the channel axis, $D_x(x)$. For this we averaged over the three bins closest to the channel axis for each channel and $x$ position (but for the top channel the total number of bins in the $y$ direction was even so we averaged over the two closest bins). The data for the central channel are shown in Fig.~\ref{fig:Diffusion_Dx_and_Dy}(b). The diffusion coefficients showed an approximately constant value in the bulk followed by an extended transition region in which it decreased toward a plateau of lower diffusivity inside the channel. We evaluated $D_x(x)$ only for $x\in[-4~\mu m,4~\mu m]$ to avoid edge effects of the finite tracking region (see \cite{DettmerRSI2014} for more details).

%%%%%%%%%%%%%%%%%%%%%%%%%%%%%%%%%%%%%%%%%%%%%%%%%

\subsubsection{Diffusion coefficient perpendicular to the channel axis}
\begin{figure}[hbtp]
	\centering
		\includegraphics[width=0.48\textwidth]{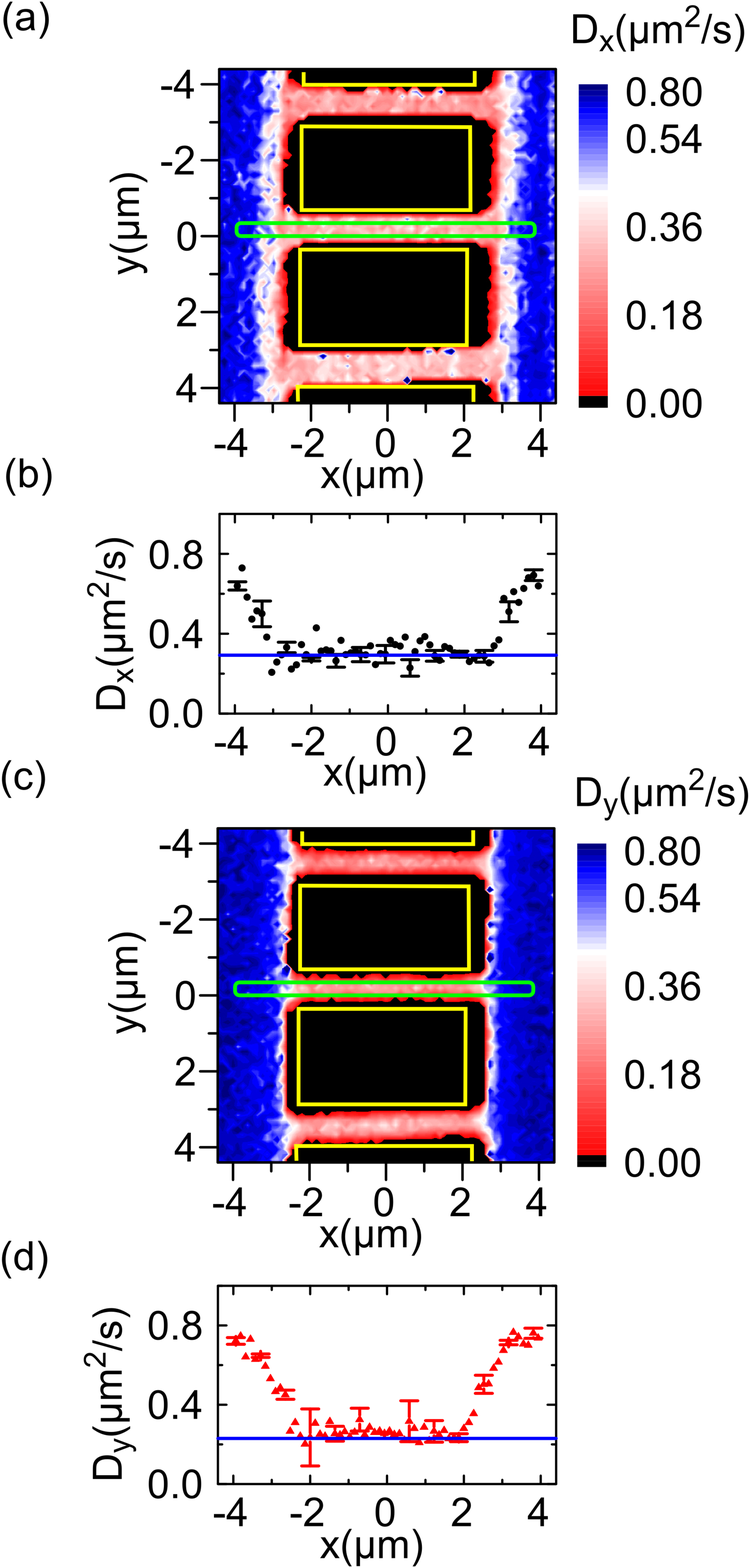}
	\caption{(Color online) Local diffusion coefficients. (a) and (c) The position-dependent diffusion coefficients presented in a color map for diffusivity parallel ($D_x$) and perpendicular ($D_y$) to the channel axis, respectively. The channel edges are marked by the yellow lines. The channels appear longer and thinner in the color map because it is based on the center position of finite size spheres. (b) and (d) The diffusivity dependence along the channel axis calculated in the marked box (green lines) by averaging over the three bins for each $x$ value for parallel [$D_x(x)$], and perpendicular [$D_y(x)$] diffusivity, respectively. Error bars are the standard deviation in between bins. For clarity error bars are only shown for points spaced every 650~nm.}
	\label{fig:Diffusion_Dx_and_Dy}
\end{figure}

\begin{figure*}[hbtp]
	\centering
		\includegraphics[width=1.0\textwidth]{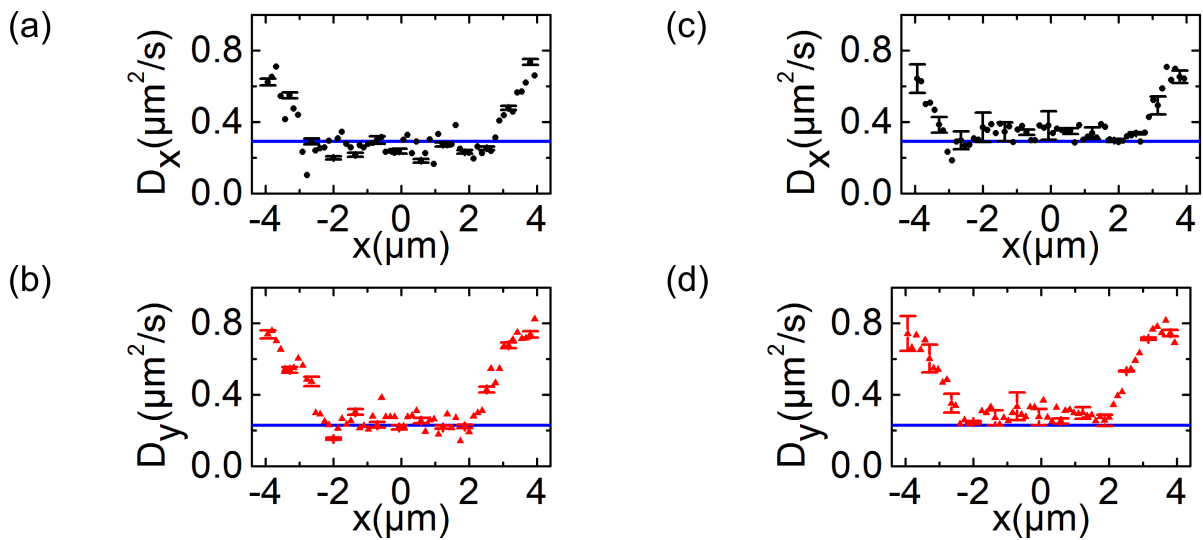}
	\caption{(Color online) Diffusion coefficient dependence along the channel axis [$D_{x,y}(x)$] for the top [(a) and (b)] and bottom [(c) and (d)] channels computed as for Fig.~\ref{fig:Diffusion_Dx_and_Dy}. Error bars are the standard deviation in between bins. For clarity error bars are only shown every 650~nm. The parallel diffusivity, $D_x(x)$, is shown in (a) and (c) and the perpendicular one, $D_y(x)$, in (b) and (d). The straight blue lines represent the values from the numerical simulations for infinitely long channels.}
	\label{fig:D_of_x_other_two_channels}
\end{figure*}

With our data we can not only quantify the parallel diffusion coefficient but also investigate diffusion perpendicular to the channel axis ($D_y$). The 2D color map of $D_y$ is shown in Fig.~\ref{fig:Diffusion_Dx_and_Dy}(c) and the dependence along the channel axis, $D_y(x)$, is shown in Fig.~\ref{fig:Diffusion_Dx_and_Dy}(d). The data for the top and bottom channels are shown in Fig.~\ref{fig:D_of_x_other_two_channels} for both parallel and perpendicular diffusivity.\newline
Within measurement accuracy, the average diffusivities inside the channels, $D_{x,ch}$ and $D_{y,ch}$, were the same for all three channels studied and agreed with our simulation values for infinitely long channels. The detailed values can be found in Table~\ref{tab:ChannelD}. The length of the transition region between bulk and channel was around ${1-1.5~\mu m}$, without significant differences between channels.  
For perpendicular diffusivity, however, the plateau inside the channel was slightly shorter ($\approx0.5~\mu m$, i.e., one particle diameter) than that of the parallel diffusivity $D_x$. Indeed, in order to reduce $D_y$, the particle has to be fully enveloped by the channel. This explains the small difference in transition length scales for $D_x$ and $D_y$. Furthermore, $D_{y,ch}$ was lower than $D_{x,ch}$ due to the motion perpendicular to the channel walls being more strongly confined than that in the parallel direction.

%%%%%%%%%%%%%%%%%%%%%%%%%%%%%%%%%%%%%%%%%%%%%%%%%%%%%%%%%%%%%%%%%%%%%%%%%%%%%%%%%%%%%%%%%%%%%%%%%%%%%%%%
We noticed that the diffusivity in both $x$  and $y$ directions in our bulk reservoirs reached a value of $D_{xy}=(0.74\pm0.06)~\mu m^2/s$ rather than the Einstein-Stokes value of $D_0=1.08~\mu m^2/s$. This can be attributed to the hydrodynamic friction exerted by the glass slide. Using Goldman's theory~\cite{Goldman1967} we estimated the average hydrodynamic separation $z$ between the particle centers and the glass surface~\cite{Bevan2000}. Inverting the theoretical relationship for the diffusivity parallel to a plane wall, $D_{xy}(a/z)/D_0$ (by using a series expansion from Happel~\cite{Happel1983}), yielded an average distance of $z=(370\pm 30)~nm$.
\begin{table}[hbtp]
	\centering
		\begin{tabular}{|c|c|c|}
	\hline	\multirow{2}{*}{{\huge $D_x$}}	& Experimental & Simulation   \\ 
	&$D_{x,ch}~(\mu m^2/s)$ & $D_{x,ch}~(\mu m^2/s)$ \\ \hline
			top channel & $0.27\pm0.05$ & $0.293\pm0.002$  \\ \hline
			central channel & $0.32\pm0.04$ & $0.293\pm0.002$    \\ \hline
			bottom channel & $0.31\pm0.03$ & $0.293\pm0.002$  \\ \hline
		\end{tabular}
	\centering
		\begin{tabular}{|c|c|c|}
\hline		\multirow{2}{*}{{\huge $D_y$}} & Experimental & Simulation   \\ 
	&$D_{y,ch}~(\mu m^2/s)$ & $D_{y,ch}~(\mu m^2/s)$ \\ \hline
			top channel &   $0.24\pm0.05$& $0.23\pm0.04$ \\ \hline
			central channel &  $0.26\pm0.03$& $0.23\pm0.04$    \\ \hline
			bottom channel &  $0.29\pm0.03$& $0.23\pm0.04$\\ \hline
		\end{tabular}
	\caption{Average parallel and perpendicular diffusion coefficients ($D_x$ and $D_y$) for all channel interiors. Experimental values were averaged from the $D_{x,y}(x)$ curves in the plateau region between $x=-2~\mu m$ and $x=+2~\mu m$. Given are the average values and the standard deviation between the different points along the $x$ axis. The simulation values were averaged from the cross-sectional values $D_{x,y}(y,z)$ over the same $y$  and $z$ ranges as the experimental ones. The simulation errors given are the standard errors of the mean over the different sampling points.}
	\label{tab:ChannelD}
\end{table}
%%%%%%%%%%%%%%%%%%%%%%%%%%%%%%%%%%%%%%%%%%%%%%%%%%%%%%%%%%%%%%%%%%%%%%%%%%%%%%%%%%%%%%%%%%%%%%%%%%
%%%%%%%%%%%%%%%%%%%%%%%%%%%%%%%%%%%%%%%%%%%%%%%%%%%%%%%%%%%%%%%%%%%%%%%%%%%%%%%%%%%%%%%%%%%%%%%%%%%%%%%%%%%%%%
\FloatBarrier
\subsection{Dependence of diffusivity on the distance from the channel axis}
\begin{figure}[htp]
	\centering
		\includegraphics[width=7cm]{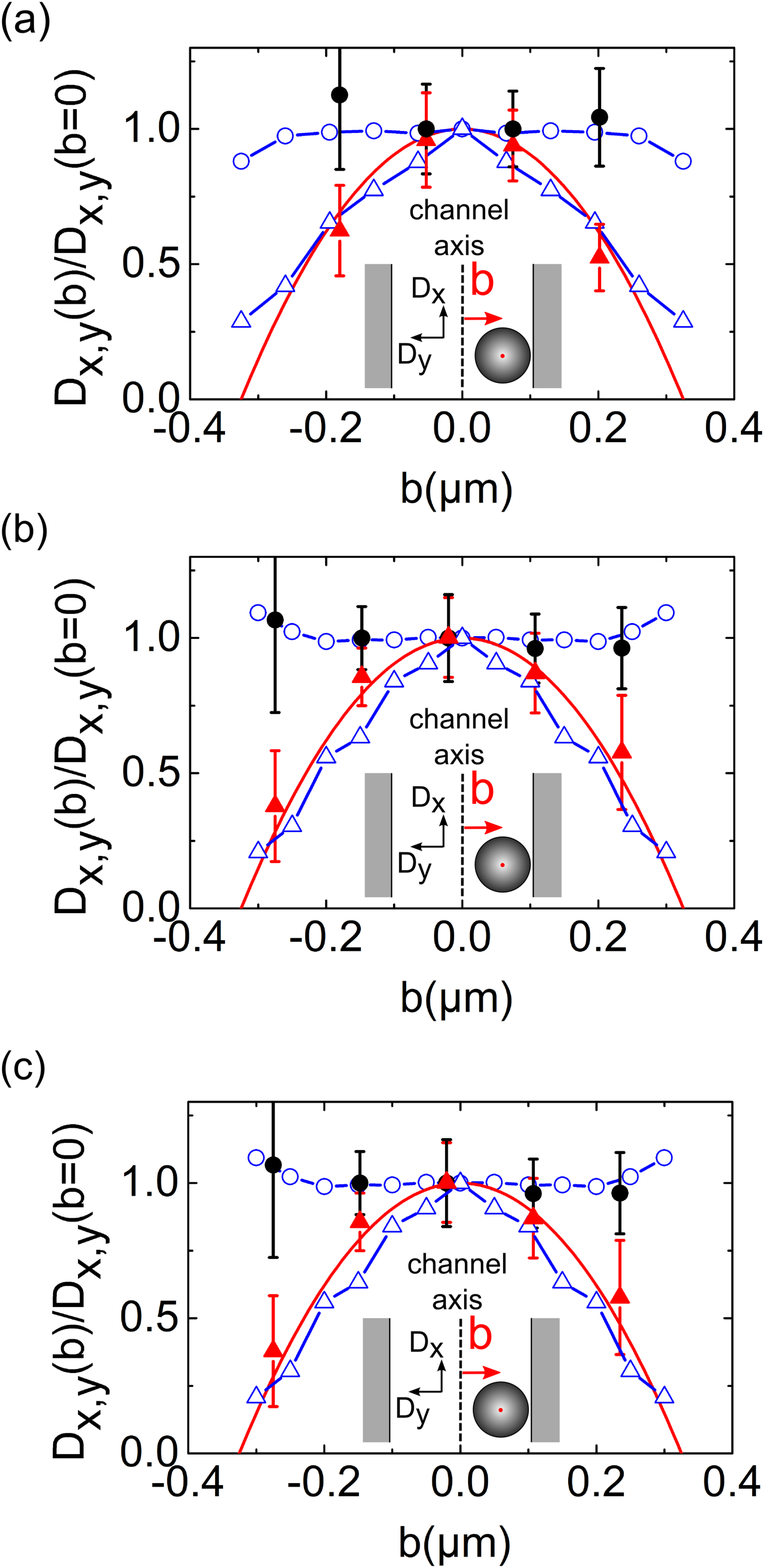}
	\caption{(Color online) Diffusion coefficient dependence along the channel width for the top (a), central (b), and bottom (c) channels. The filled circles $\bullet$ are the measured diffusion coefficients parallel to the channel axis ($D_x$) and the filled triangles $\blacktriangle$ are the ones in the perpendicular direction ($D_y$). The error bars are the standard deviations between the different bins that were averaged over.	The lines with empty symbols are values from the numerical simulations for $D_x$ ($\circ$) and $D_y$ ($\triangle$). The continuous lines show the empirical parabolic dependence according to Eq.~(\ref{eq:empirical_dependence}). The inset illustrates the definition of the coordinate system.}
	\label{fig:D_of_y}
\end{figure}

For the channel interior, we calculated the dependence of the diffusivity on the distance $b$ from the channel axis, $D_x(b)$ and $D_y(b)$. We averaged over all bins between $x=-2~\mu m$ and $x=+2~\mu m$ for each $y$ value up to the channel walls. The data for $D_y$ ~(triangles in Fig.~\ref{fig:D_of_y}) show that the diffusion coefficient is at a maximum in the channel center. As the particle is moving closer to the channel wall, $D_y$ drops significantly, as expected when the particle approaches the channels walls.
In stark contrast, the diffusivity parallel to the channel axis ($D_x$) remained almost constant throughout the entire channel width~(circles in Fig.~\ref{fig:D_of_y}). This is contrary to expectations based on hindered diffusion in proximity to plane walls. We observed the same dependence for all three channels.
Empirically, the perpendicular diffusivity was reasonably well described by the parabolic equation:
\begin{equation}
\frac{D_y(b)}{D_y(0)}\approxeq 1-\left( \frac {b}{w/2-a} \right)^2
\label{eq:empirical_dependence}
\end{equation}
where $w$ is the width of the channel and $a$ the particle radius. 
Due to the complex geometric shape of the channel, we expect that no closed analytical form for the dependence of $D_y(b)$ exists. However, the surprising agreement of the data with the parabolic equation~(\ref{eq:empirical_dependence}) suggests that the dependence can be treated successfully in low orders of a perturbative expansion. In that sense Eq.~(\ref{eq:empirical_dependence}) represents an expansion up to second order of the true relationship. 
By fitting this empirical relationship to the data we determined the position of the channel axis~($b=0$) from the maximum of the parabola at sub pixel resolution as well as the on-axis diffusivity $D_y(b=0)$. This was important since this allowed us to define the parallel on-axis diffusivity $D_x(b=0)$ as the measured $D_x(b)$ value closest to the center for an uneven number of bins in the $y$ direction or the average of the two closest bins in the case of an even number of bins.

To our knowledge, the surprising behavior of $D_x(b)$ that we found here has not previously been observed experimentally in microfluidic channels. Only analytical and numerical studies on the hydrodynamic drag force experienced by spherical particles translating in closely fitting cylindrical channels~\cite{Bungay1973,Bhattacharya2010} have predicted this kind of dependence. Thus, our experiments allowed for the first qualitative experimental testing of their predictions in very close confinements on the submicron scale. Despite the lack of an analytical solution due to the cross sections of our channels being semi elliptical rather than cylindrical we could compare our measurements to our finite element simulations and found good agreement (with absolute values for $D_x$ and $D_y$ being shown in Table~\ref{tab:ChannelD} and the $b$ dependence in Fig.~\ref{fig:D_of_y}). This comparison shows that hindered diffusion behaves qualitatively differently in closely confining channels as compared to more extended geometries due to hydrodynamic interactions determined by the microchannel geometry. We try to give an intuitive explanation of this phenomenon in Appendix~\ref{sec:tentative_explanation}.

%%%%%%%%%%%%%%%%%%%%%%%%%%%%%%%%%%%%%%%%%%%%%%%%%%%%%%%%%%%%%%%%%%%%%%%%%%%%%%%%%%%%%%%%%%%%%%%%%%%%%%%%

%%%%%%%%%%%%%%%%%%%%%%%%%%%%%%%%%%%%%%%%%%%%%%%%
\subsection{Simulations for the full cross-sectional channel profile and additional particle-channel size ratios}
\begin{figure}[hbtp]
	\centering
		\includegraphics[width=8.5cm]{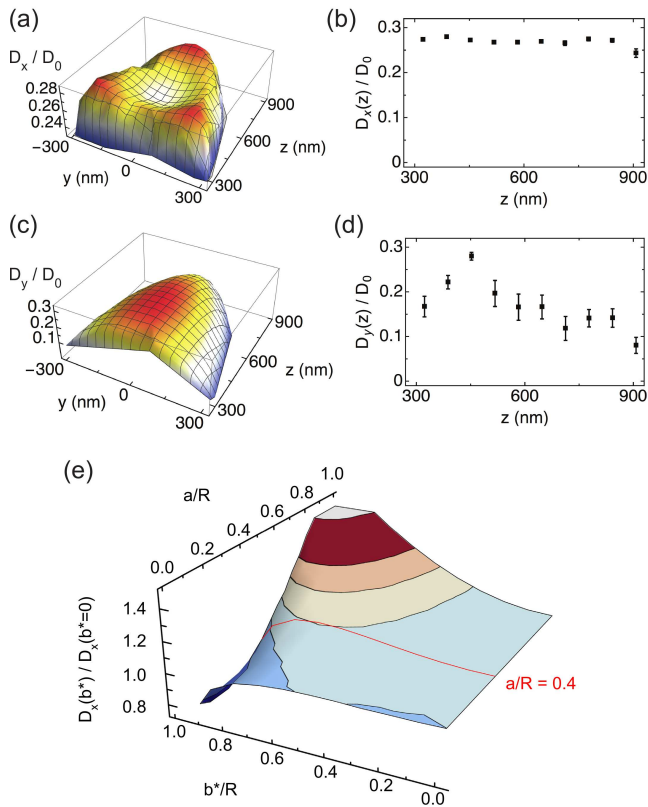}
	\caption{(Color online) Finite element simulations for the diffusivity dependence across the full cross-sectional channel profile and additional particle-channel-size ratios. (a) and (c) The 3D landscape showing the cross-sectional profile of the parallel (a) and perpendicular (c) diffusivity of 500-nm spheres confined by an infinitely long semi elliptical channel. The diffusivity values were normalized by the Stokes-Einstein value $D_0$. The coordinates refer to the center position of the finite size spheres. The roughness of the plots is due to random position sampling. (b) and (d) The diffusivity dependence on the particle elevation $z$ for parallel [$D_x(z)$] and perpendicular diffusivity ($D_y(z)$) is obtained by binning the data from (a) and (c). (e) The 3D landscape showing the parallel diffusivity of spheres of various sizes (radius $a$) confined by infinitely long cylindrical channels (radius $R$) at different normalized off-axis displacements [$b^{*}=b/(R-a)$].}
	\label{fig:additional_simulations}
\end{figure}

Our finite element simulations covering the entire channel cross section of an infinitely long semi elliptical channel  [Figs.~\ref{fig:additional_simulations} (a) and ~\ref{fig:additional_simulations}(c)] show that this interesting effect is not strongly dependent on the average elevation of the particle in the channel but occurs across the entire channel height. The parallel diffusivity ($D_x$) showed only small variations across the cross section: from $\sim 0.25 D_0$ to $\sim 0.28D_0$. The qualitative features of our simulated $D_x$ agree well with classical analytical solutions of diffusion in circular channels~\cite{Deen1987}. On the other hand, the perpendicular diffusivity ($D_y$) varied more strongly: from almost zero to $\sim 0.32 D_0$. Close to the walls we observed a drop in diffusivity for both $D_x$ and $D_y$ due to the expected rapid increase in friction exerted by the channel walls. 
The dependence of diffusivity on elevation [Figs.~\ref{fig:additional_simulations}(b) and ~\ref{fig:additional_simulations}(d)] was rather flat for $D_x$. The perpendicular diffusivity ($D_y$) on the other hand showed a non uniform dependence on elevation with a peak corresponding to the furthest distance from the walls. This behavior is qualitatively similar to that observed for lateral displacements (see Fig.~\ref{fig:D_of_y}). \newline
Furthermore, we investigated the influence of the ratio of particle to channel size on the diffusivity profile $D_x(y)$. To this end we performed finite element simulations for particles in infinitely long cylindrical channels (a cylindrical geometry being chosen for greater computational efficiency) [Fig.~\ref{fig:additional_simulations}(e)]. Here we defined the ratio of particle to channel radius, $a/R$, and the normalized off-axis displacement of the particle relative to the channel radius, $b^{*}/R=b/(R-a)R$, where  $b^{*}/R = 0$ corresponds to a particle at the center of the channel and $b^{*}/R = 1$ to a particle touching the channels walls. 
For small size ratios $a/R$ we recover the well-known monotonic decrease of diffusivity toward the walls resembling Fax\'en's law. Indeed, for small particle sizes, the colloid ``effectively sees'' a flat wall. At larger ratios the confinement of the channel and the curved boundary become important and the diffusivity profile flattens out to approximately constant diffusivity across the entire channel. This is the case relevant to our experimental study. At even larger ratios the profile expected from Fax\'en's law even gets reversed and the diffusivity increases in proximity to the walls.

%%%%%%%%%%%%%%%%%%%%%%%%%%%%%%%%%%%%%%%%%%%%%%%%%%%%%%%%%%%%%%%%%%%%%%%%%%%%
%%%%%%%%%%%%%%%%%%%%%%%%%%%%%%%%%%%%%%%%%%%%%%%%%%%%%%%%%%%%%%%%%%%%%%%%%%%
\section{Conclusions}
In summary, we presented the detailed measurement of the position-dependent diffusion coefficients of spherical particles closely confined by finite length channels in directions parallel and perpendicular to the channel axis. Of particular interest to models of channel-facilitated diffusion is the determination of the dependence along the channel axis, $D_x(x)$, showing that diffusion in the channel interior behaves as if the channels were infinitely long with an almost constant diffusivity throughout the entire channel length.\newline
Furthermore, we observed the parallel diffusivity to remain approximately constant throughout the entire channel width, in contrast to the perpendicular diffusivity that decreased toward the channel walls.\newline
We expect that our findings will stimulate further studies of the special features of Brownian motion arising in strong confinement, which is commonplace in cellular environments. Besides this potential for exciting new insights into biophysics, the physical process of confined Brownian motion is strongly linked to low-Reynolds-number hydrodynamics in closely confining environments as governing flow in the thriving fields of micro- and nanofluidics. Our results could be of interest to efficiently control particle transport in technological applications such as, e.g., the construction of drift ratchets for particle sorting.
%%%%%%%%%%%%%%%%%%%%%%%%%%%%%%%%%%%%%%%%%%%%%%%%%%%%%%%%%%%%%%%%
%%%%%%%%%%%%%%%%%%%%%%%%%%%%%%%%%%%%%%%%%%%%%%%%%%%%%%%%%%%%%%%%%%%%%%%%%%%
\begin{acknowledgments}
S.L.D. acknowledges funding from the German Academic Exchange Service~(DAAD) and the German National Academic Foundation. S.P. and U.F.K. were supported by an ERC starting grant. S.P. also acknowledges support from the Leverhulme Trust and the Newton Trust through an Early Career Fellowship. We thank Sandip Ghosal for helpful discussions. 
\end{acknowledgments}

\appendix
\section{STOKES-EINSTEIN RELATION}
\label{sec:Stokes-Einstein}
The Stokes-Einstein relation~\cite{Einstein1905} is given by ${\mathbf{D}=k_BT\boldsymbol{\nu}^{-1}}$, with the Boltzmann constant $k_B$, absolute temperature $T$, and the viscous friction tensor $\boldsymbol{\nu}$. The viscous friction tensor relates the hydrodynamic drag force $\vec{F}$ to the velocity $\vec{v}$ of a particle translating in a quiescent fluid: $\vec{F}=\boldsymbol{\nu}\vec{v}$. For a sphere suspended in an infinite fluid of (temperature-dependent) viscosity $\eta(T)$ this is $\boldsymbol{\nu}_0=\nu_0 \openone,\nu_0=6\pi a \eta(T)$, leading to the Stokes-Einstein diffusivity 
\begin{equation}
\mathbf{D}=D_0\openone,D_0=k_BT/\nu_0=k_BT/6\pi a \eta(T).
\label{eq:Stokes-Einstein-D}
\end{equation}
In hindered diffusion, the friction and diffusion coefficients are no longer the same for the different axial directions and they become position dependent~\cite{Happel1983,Faucheux1994}. In two dimensions we have
\begin{equation}
\boldsymbol{\nu}=\boldsymbol{\nu}(\vec{r})=\begin{pmatrix} \nu_x(\vec{r}) & 0 \\ 0 & \nu_y(\vec{r}) \end{pmatrix},
\label{eq:hindered_drag_matrix}
\end{equation}
\begin{align}
\mathbf{D}=\mathbf{D}(\vec{r})&=\begin{pmatrix} D_x(\vec{r}) & 0 \\ 0 & D_y(\vec{r}) \end{pmatrix} \notag \\
%&=D_0 \begin{pmatrix} D_x(\vec{r})/D_0 & 0 \\ 0 & D_y(\vec{r})/D_0 \end{pmatrix} \notag \\
&=D_0 \begin{pmatrix} \nu_0/\nu_x(\vec{r}) & 0 \\ 0 & \nu_0/\nu_y(\vec{r}) \end{pmatrix},
\label{eq:hindered_diffusivity_matrix}
\end{align}
which connects position-dependent viscous friction to hindered diffusion coefficients. The friction coefficients were then calculated by numerically solving the Stokes equation with the finite element method~(COMSOL Multiphysics 4.3b with creeping flow module). We thus arrived at the ratios of friction coefficients $\nu_0/\nu_x(y,z)$ and $\nu_0/\nu_y(y,z)$, which, by Eq.~(\ref{eq:hindered_diffusivity_matrix}), together with $D_0$, give the diffusion coefficients $D_x(y,z)$ and $D_x(y,z)$.
%%%%%%%%%%%%%%%%%%%%%%%%%%%%%%%%%%%%%%%%%%%%%%%%%%%%%%%%%%%%%%%%%%%%%%%%%%%%%%%%%%%%%%
%%%%%%%%%%%%%%%%%%%%%%%%%%%%%%%%%%%%%%%%%%%%%%%%%%%%%%%%%%%%%%%%%%%%%%%%%%%%%%%%%%%%%
\section{TENTATIVE EXPLANATION OF CONSTANT PARALLEL DIFFUSIVITY ACROSS THE CHANNEL WIDTH}
\label{sec:tentative_explanation}
%\newcounter{appendixfigure}
%\setcounter{appendixfigure}{1}
%\renewcommand{\thefigure}{B.\arabic{appendixfigure}}

\begin{figure}[hbtp]
	\centering
		\includegraphics[width=0.50\textwidth]{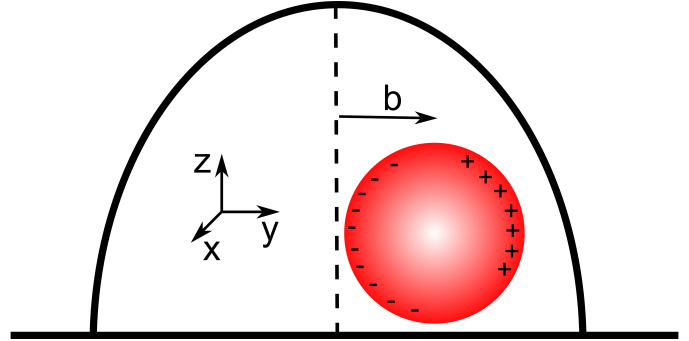}
	\caption{(Color online) Tentative explanation for constant parallel diffusivity across the channel width. Drag increase ($+$ signs) and decrease ($-$ signs) on opposite sides of the spherical particles balance each other due to the close confinement.}
	\label{fig:Drag_decrease}
\end{figure}

A graphical illustration of the tentative explanation of the constant parallel diffusivity across the channel width is presented in Fig.~\ref{fig:Drag_decrease}. While there is an increased drag force on the side of the sphere approaching the channel wall, due to the close confinement, the opposite side of the sphere moves away from the other wall and experiences a decreased drag. We assume that the increase in drag at a single point will be greater than the decrease on the other side but this gets balanced by a larger surface area opposite to the approaching side. For the diffusivity perpendicular to the channel walls, the drag increase is a lot steeper than for the parallel diffusivity, as is regularly observed in proximity to close walls. Therefore the larger surface area can no longer balance the drag increase and the total diffusivity decreases.

%%%%%%%%%%%%%%%%%%%%%%%%%%%%%%%%%%%%%%%%%%%%%%%%%%%%%%%%%%%%%%%%%%%%%%%%%%%%%%%%%%%%%%
%%%%%%%%%%%%%%%%%%%%%%%%%%%%%%%%%%%%%%%%%%%%%%%%%%%%%%%%%%%%%%%%%%%%%%%%%%%%%%%%%%%%%

\FloatBarrier
%\bibliography{library}

%merlin.mbs apsrev4-1.bst 2010-07-25 4.21a (PWD, AO, DPC) hacked
%Control: key (0)
%Control: author (8) initials jnrlst
%Control: editor formatted (1) identically to author
%Control: production of article title (-1) disabled
%Control: page (0) single
%Control: year (1) truncated
%Control: production of eprint (0) enabled
%

\end{document}